\documentclass[9pt,twocolumn,twoside]{opticajnl}
\journal{opticajournal} 

\setboolean{shortarticle}{false}

\usepackage{lineno}

\title{Hybrid biphoton spectrometer for time-resolved quantum spectroscopy across visible and near-infrared regions}

\author[1,$\dagger$,$\ddagger$]{Ozora Iso}
\author[1,$\dagger$]{Koya Onoda}
\author[1,$\dagger$]{Nicola J. Fairbairn}
\author[2]{Masahiro Yabuno}
\author[2]{Hirotaka Terai}
\author[2]{Shigehito Miki}
\author[1,3,*]{Ryosuke Shimizu}

\affil[1]{Department of Engineering Science, Graduate School of Informatics and Engineering, The University of Electro-Communications, 1-5-1 Chofugaoka, Chofu, 182-8585, Tokyo, Japan.}
\affil[2]{Advanced ICT Research Institute, National Institute of Information and Communications Technology, 588-2 Iwaoka, Nishi-ku, Kobe, 651-2492, Hyogo, Japan.}
\affil[3]{Institute for Advanced Science, The University of Electro-Communications, 1-5-1 Chofugaoka, Chofu, 182-8585, Tokyo, Japan.}

\affil[$\dagger$]{These authors contributed equally to this work.}
\affil[$\ddagger$]{Present address: School of Chemistry, University of Glasgow, Glasgow, G12 8QQ, United Kingdom.}
\affil[*]{r-simizu@uec.ac.jp}

\begin{abstract}
Joint spectral measurements are a powerful tool for characterising biphoton spectral correlation, which is crucial for quantum information and communication technologies. In these applications, highly pure biphoton states are essential in any time- and frequency-mode, often obviating the need for time-resolved measurements. Conversely, spectroscopy utilising entangled photon pairs is gaining significant attention for its ability to unveil molecular dynamics, a field that critically demands time-resolved capabilities. Here, we introduce a methodology for capturing a biphoton spectrum that comprises visible and near-infrared photons, resulting in a highly non-degenerate joint spectrum. Our system employs two non-scanning spectrographs: a fibre spectrometer for near-infrared photons and a delay-line-anode single-photon imager for visible photons. We successfully measure the joint spectral intensity by leveraging a time-tagging acquisition strategy. Furthermore, our approach uniquely enables time-resolved joint spectral measurements with respect to the laser synchronisation against the hundreds-of-picosecond instrument response function. Our methodology could advance heralded fluorescence spectroscopy using biphoton sources to investigate temporal dynamics of complex biological, chemical, and physical systems.
\end{abstract}

\setboolean{displaycopyright}{false} 

\begin{document}

\maketitle

\section{Introduction}
\begin{figure*}[!t]
\begin{center}
\includegraphics[width=1\linewidth]{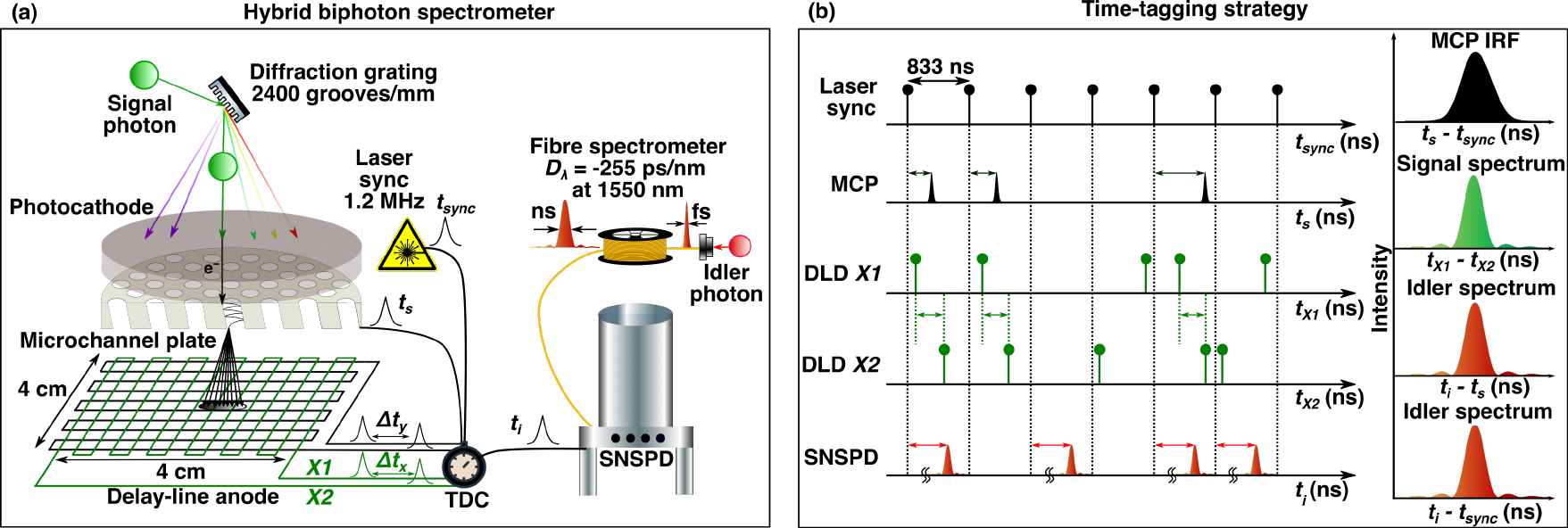}
\caption{(a) The principle of the hybrid biphoton spectrometer. The signal photons are measured with the delay-line-anode single-photon detector (DLD), which consists of the photocathode, the microchannel plate (MCP), and the delay-line anode. The idler photons are temporally stretched by the fibre spectrometer, followed by a superconducting nanowire single-photon detector (SNSPD). The joint spectral intensity (JSI) is time-resolved according to the laser synchronisation signal (laser sync). (b) The time-tagging strategy illustrates how electrical pulses are processed at the Time-to-digital Converter (TDC). All data processing is based on the second-order correlation function, or time-difference measurement. The four one-dimensional histograms on the right are used to reconstruct a time-resolved JSI. }
 \label{Main_Figure1}
 \vspace{-7mm}
 \end{center}
\end{figure*}
The exploitation of quantum entanglement in photon pairs has opened up new frontiers in various photonic research areas, from quantum information processing to emerging applications in biophotonics and chemical physics. In quantum technology, entangled photons are one of the possible resources for developing quantum computers \cite{ladd2010quantum}, secure communication networks \cite{gisin2007quantum}, and metrology tools \cite{toth2014quantum}. Concurrently, a new paradigm is emerging in chemical and biophotonics, where entangled photons are used as a non-classical light source to probe molecular dynamics \cite{li2023single}. This approach offers unique advantages over traditional classical spectroscopy, particularly through the use of time- and frequency-mode correlations of biphotons \cite{alvarez2025entangled}.
For this emerging field to mature, advanced spectroscopic techniques capable of capturing the subtle, fast dynamics of molecular systems are required. 
\newline
\indent
Characterising photon pairs in the frequency-time degrees of freedom is important for applications in the quantum information community, enabling encoding methods such as time-bin or frequency-bin measurements. The joint temporal intensity is a method for characterising the temporal correlations of entangled photon pairs using ultrafast detection techniques \cite{maclean2018direct}, whereas its conjugate, the joint spectral intensity (JSI), can be measured to characterise the frequency correlations \cite{Zielnicki07062018}. However, JSI's static nature fundamentally limits its application in time-sensitive studies. Consequently, a key challenge and a highly appealing goal in quantum optics is the simultaneous measurement of the complete two-photon waveform in both the time and frequency domains \cite{jin2018time, davis2020measuring, Kuwana_2026}, which is essential for fully evaluating and tailoring entangled wave packets. Current JSI measurement techniques - which include scanning dispersive optics \cite{PhysRevA.80.033814}, dual fibre spectroscopy \cite{avenhaus2009fiber}, stimulated emission tomography \cite{Fang:14}, and single-photon-sensitive cameras \cite{lubin2021heralded} - have made significant strides in efficiency. Yet, none can provide the picosecond temporal resolution necessary to track fast molecular processes. This gap highlights a critical need for a new class of instrumentation: a time-resolved biphoton spectrometer.
\newline
\indent
We address this need by presenting the design, construction, and validation of a novel picosecond time-resolved hybrid biphoton spectroscopic system. The system combines two distinct, highly sensitive single-photon spectrographs: a fibre-based spectrograph \cite{avenhaus2009fiber} employing a superconducting nanowire single-photon detector (SNSPD) for near-infrared photons, and a delay-line-anode single-photon detector \cite{czasch2007position} for visible photons. 
Our system is first validated by generating and measuring the highly non-degenerate photon pairs from a lithium triborate (LBO) crystal. We then present a comprehensive analysis, including one-dimensional spectra of the signal and idler photons, a comparison of the simulated JSI with experimentally reconstructed data, and a demonstration of a JSI resolved on a picosecond timescale. 
\newline
\indent
While the absolute spectra of signal and idler photons are independently measured, the system requires only the relative arrival time between detectors to measure time-resolved JSI. Furthermore, time-resolved JSI can be obtained by measuring only the temporal characteristics of the signal photons, since time-tag analysis can extract coincidence events only when signal photons are detected at times separated by the photodiode synchronisation. The absolute timing or phase is irrelevant for our analysis. This uniqueness arises from the flexibility of the Time-to-digital converter (TDC) program, in which time tags from three different detectors are correlated and temporal filtering is implemented.  
\newline
\indent
\begin{figure*}[!t]
\begin{center}
\includegraphics[width=1\linewidth]{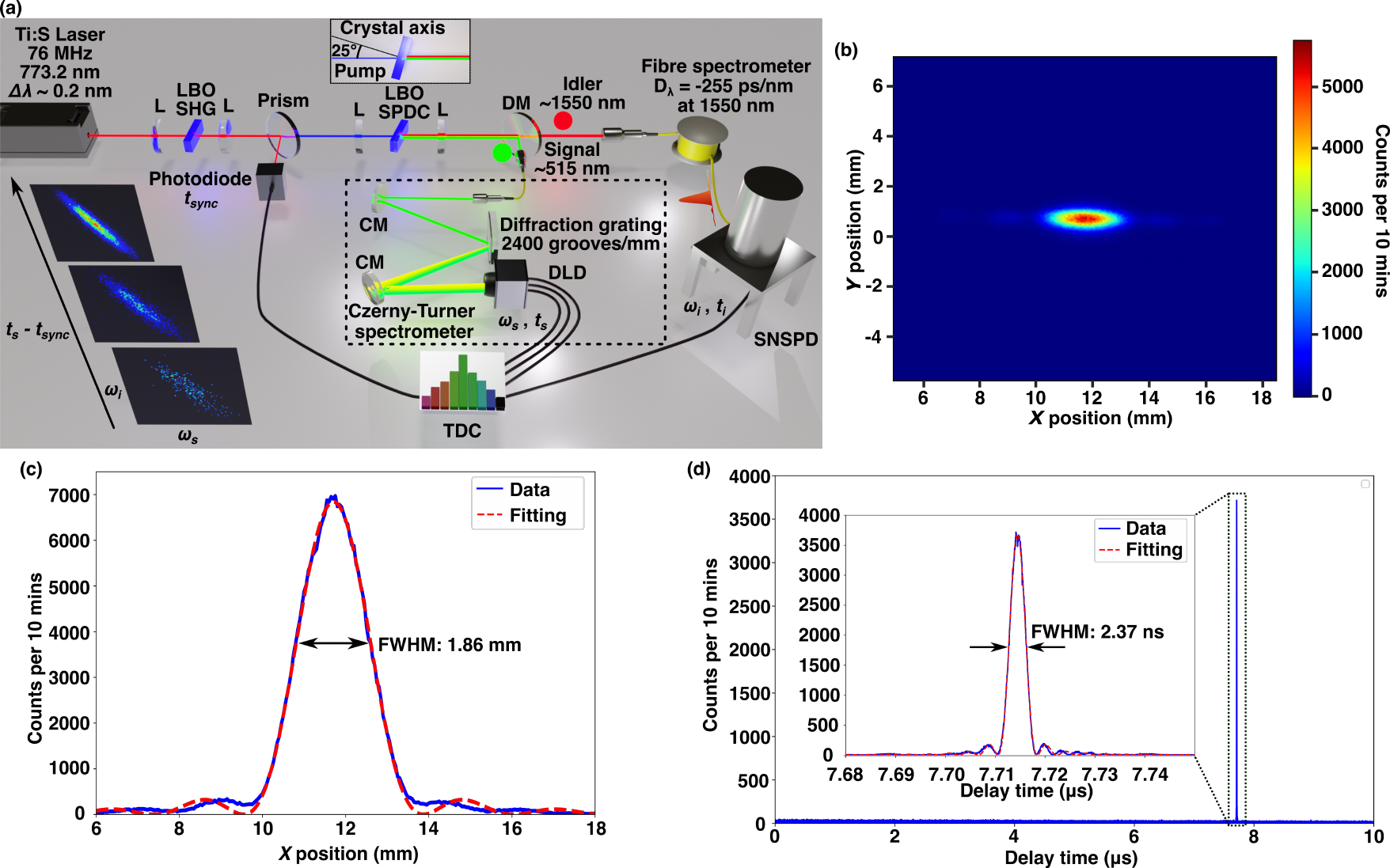}
\caption{(a) Optical diagram showing the time-resolved detection of highly non-degenerate photon pairs using the hybrid biphoton spectrometer. The top inset shows the phase-matching angle between the excitation laser path and the crystal axis of the Lithium triborate (LBO) crystal used for spontaneous parametric down-conversion (SPDC). The phase-matching angle is $25^{\circ}$. The bottom left figures show typical examples of time-resolved JSI, where the horizontal and vertical axes represent the signal ($\omega_{s}$) and idler ($\omega_{i}$) frequencies, respectively. L: Lens, SHG: Second-harmonic generation, DM: Dichroic mirror, CM: Concave mirror, SNSPD: Superconducting nanowire single-photon detector, TDC: Time-to-digital converter. (b) shows the captured image at the DLD, and its projection is (c), corresponding to the spectrum of signal photons. (d) shows a coincidence peak between the MCP and SNSPD. The inset is an enlarged histogram providing the spectrum of idler photons.}
 \label{Main_Figure2}
 \vspace{-7mm}
 \end{center}
\end{figure*}
Figure \ref{Main_Figure1}(a) depicts the principle of the hybrid spectrometer. The delay-line-anode single-photon detector (DLD, RoentDek, DLD40) measures the visible signal photon ($\sim$515 nm) spectrally resolved by a 2400 grooves/mm grating spectrometer. DLD is a position-sensitive detector based on the photoelectric effect, and it is generally applied for X-ray \cite{li2025imaging}, ultraviolet \cite{schnorr2014electron}, and electron imaging \cite{gong2022attosecond}. However, its potential for photon-pair detection is not fully exploited in the quantum optical context. We previously demonstrated photon-pair detection generated by a copper chloride semiconductor crystal in the near-ultraviolet range \cite{Iso:25}, yet a demonstration using spontaneous parametric down-conversion (SPDC) has not been achieved.
Although the DLD principle is described in our previous work, we briefly summarise it here. DLD consists of the photocathode, the microchannel plate (MCP), and the meander-wired delay-line-anode sensor. An impinged photon is converted into a photoelectron and amplified via the MCP, then electrically detected at the end of the sensor \cite{JAGUTZKI2002244}. The electrical signals split in opposite directions from the detection point, and the time-of-arrival difference between them is recorded. Due to the one-to-one correspondence of the photon arrival position and the time difference, its position is calculated as
\begin{eqnarray}
\label{DLD_position_estimation}
x (\Delta t_{x}) &=& \frac{(\Delta t_{x} + t_a)v}{2},\\
y (\Delta t_{y}) &=& \frac{(\Delta t_{y} + t_a)v}{2}, 
\end{eqnarray}
where $\Delta t_{x}$,\ $\Delta t_{y}$ is the time-difference between two electrical signals generated by $x$- and $y$-wired delay-line-anode, respectively. $t_{a}$ is the propagation time from the edge-to-edge of the delay-line anodes, $v$ is the propagation speed, $x$ and $y$ are specified photon arrival coordinates. $x$ is calculated with the wire end, denoted as $X1$ and $X2$, and used to reconstruct the spectrum. The $x$-projection of the DLD image is used to reconstruct the biphoton spectrum. 
The other spectrometer is a fibre spectrometer that detects the arrival times of colour-separated idler photons using the SNSPD. The dispersion-compensating fibre (DCF) is selected to stretch the photon wave packet from femtosecond to nanosecond widths, which are measurable using a TDC. The dispersion coefficient (denoted as $D_{\lambda}$) is $-255\:\mathrm{ps}/\mathrm{nm}$ at 1550 nm. The one-dimensional spectra from both spectrometers are integrated into two-dimensional spectral maps as a JSI. In addition to the spectral information of photon pairs, a synchronisation signal (laser sync) from a laser is provided to the TDC to time-resolve the JSI. The repetition rate of the synchronisation is reduced to 1.2 MHz by a signal divider to prevent overflow in the TDC.  
\newline
\indent
Figure \ref{Main_Figure1}(b) shows a time-tagging strategy of the hybrid spectrometer associated with the obtained one-dimensional histograms. The TDC flexibly changes to process a stream of time-tagging data from all detectors. The following acquisition is set by the dedicated software distributed by the manufacturer. The first five channels receive electrical signals from the DLD, while the sixth and seventh channels register the signal from the SNSPD and laser, respectively. The synchronisation signals appear every 833 ns, which is the reciprocal of the repetition rate of the divided signals, triggering signals for the MCP and SNSPD. The time difference between the MCP and the synchronisation signal reflects the instrument response function (IRF) of the MCP. The DLD $X1$ and DLD $X2$, corresponding to the end of the $X1$ and $X2$ wires in (a),  are used to locate the spectrum of signal photons when triggered by MCP signals. The idler photons, dispersed by the fibre, are measured by the start signals of the MCP and the laser synchronisation. It should be noted that SNSPD signals arrive after microseconds, reflected in the unit of the SNSPD time-tagging graph, because the idler photons propagate through a long fibre. 

\section{Results}\label{sec2}
\subsection{The one-dimensional spectral measurement of signal and idler photons}
\begin{figure*}[!t]
\begin{center}
\includegraphics[width=\linewidth]{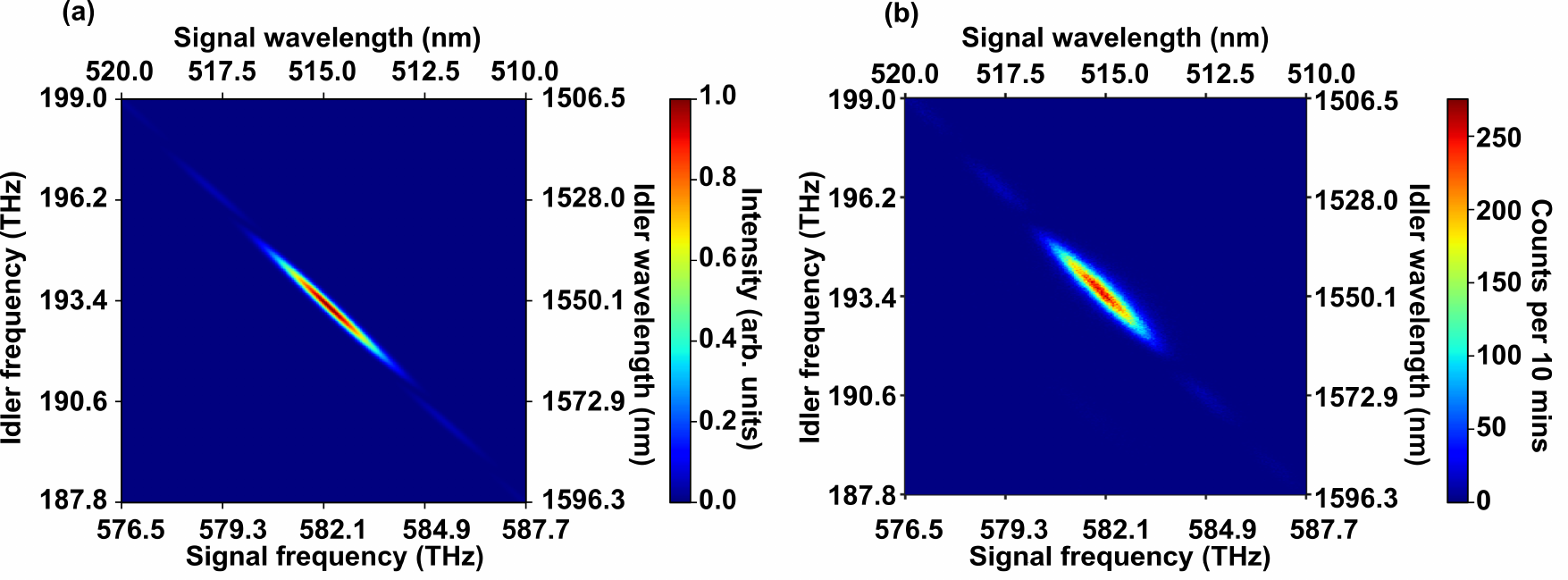}
\caption{(a) Simulated JSI of the LBO crystal, and (b) experimentally measured JSI.}
 \label{Main_Figure3}
 \end{center}
 \vspace{-7mm}
\end{figure*}

Figure \ref{Main_Figure2}(a) depicts the optical schematics of time-resolved JSI measurement of highly non-degenerate JSI with the hybrid biphoton spectrometer. The second-harmonic light is generated via the 5 mm-long LBO crystal excited by the mode-locked Ti:S laser, with a peak at 773.2 nm. The spectral linewidth is around 0.2 nm, and the laser repetition rate is 76 MHz. The excitation source is incident on another LBO crystal at a power of 52 mW, inducing SPDC to generate non-degenerate photon pairs. The inset shows the geometry of the LBO crystal before and after excitation. The phase-matching angle is calculated as $25^{\circ}$ using LBO crystal properties, and it is custom-ordered to fulfil the phase-matching function (see \href{Supplementary link}{Supplementary Information}). Two daughter photons are separated by the dichroic mirror (DM), which reflects the signal photons ($\sim$515 nm) and transmits the idler photons ($\sim$1550 nm). The idler photons are dispersed through a fibre spectrometer, followed by the SNSPD, while the signal photons are detected with the DLD installed at the output of the Czerny-Turner spectrometer. Here, the Czerny-Turner spectrometer consists of two concave mirrors and the diffraction grating with 2400 grooves/mm. The first mirror collimates the diverged beam from the single-mode fibre, and the second mirror focuses the spectrally dispersed beam onto the DLD surface. The one-dimensional spectrum is measured horizontally on the DLD sensor surface, using signals from the two ends of the delay-line anode. In parallel, the MCP inside the DLD provides temporal information that correlates with the time tags of the SNSPD and photodiode. The signal count on the DLD is $(6.3\pm 0.1)\times 10^4\: \mathrm{Hz}$, the idler count on the SNSPD is $(8.9\pm 0.1)\times 10^4\: \mathrm{Hz}$, and the coincidence count is $(2.6\pm 0.1)\times 10^4\: \mathrm{Hz}$.
\newline
\indent
Figure \ref{Main_Figure2}(b) shows the image captured by the DLD in which the horizontal ($x$) and vertical ($y$) axes are the spatial coordinates. The incident photons are horizontally dispersed by the grating, indicating the main centre peak alongside the faint side lobes. The $x$ position is projected as the one-dimensional spectrum in Fig. \ref{Main_Figure2}(c), clearly showing the central peak with continuous background counts. These unrelated counts are generated by thermal noise on the photocathode, where a high voltage is applied to induce the photoelectric effect. The spatial width of the dispersed photons is estimated to be 1.86 mm using a sinc-squared function. It is worth mentioning that the $y$-directional information is discarded in our scheme. Therefore, our system still has the capacity to add measurable parameters, such as spatial or temporal information. The typical suggestion is to utilise a fibre bundle to allocate the spatial information to the $y$ coordinate, or insert a streak tube to capture ultrafast dynamics.
Figure \ref{Main_Figure2}(d) is the output profile of colour-separated timing signals after the DCF. The coincidence peak exists at approximately $7.7\; \mu\mathrm{s}$ due to the propagation delay through the DCF. The peak is magnified, as shown in the inset, which features a distinct side lobe in the phase-matching function. The temporal width of the central peak is estimated to be 2.37 ns using a sinc-squared function.
The spectral resolution is also calculated to be 0.63 nm and 0.56 nm for the signal and idler spectrometer, respectively. The signal is estimated using spectral lines from a low-pressure mercury lamp \cite{sansonetti1996wavelengths}, and a derived reciprocal linear dispersion of 0.95 nm/mm converts spatial resolution to the spectral counterpart of the DLD. For the idler, the temporal jitter of the SNSPD is simply divided by the dispersion coefficient.
\subsection{JSI measurement}
The expected JSI is first numerically simulated based on the nonlinear crystal characteristics. In our scenario, a type-I LBO crystal is employed, which only requires angle-dependent phase-matching conditions. 
Figure \ref{Main_Figure3}(a) shows the simulated JSI, where the centre wavelengths of signal and idler photons are 582.1 THz (515.0 nm) and 193.4 THz (1550 nm), respectively. The projected frequency widths are estimated at 1.80 THz for signal photons and 1.74 THz for idler photons. 
Figure \ref{Main_Figure3}(b) shows the experimentally measured JSI. The horizontal and idler vertical wavelengths originate from individual spectra. The horizontal axis of the JSI is driven by the MCP and stopped by the SNSPD, whereas the vertical axis represents the image projection onto the DLD. The projected frequency width is 1.98 THz for signal photons and 1.87 THz for idler photons, respectively.
\newline
\indent
A clear discrepancy exists regarding the projected frequency width as well as diagonal spread of the peak between simulated and experimental data. The major distortion comes from the detector's resolution. The effect is decomposed into the signal and idler spectra, each of which affects the spectral spread of the horizontal and vertical directions of the JSI. The signal spatial distortion originates from the timing jitter of $\Delta t_{x}$ in Eq. (\ref{DLD_position_estimation}). The idler spectral spread is attributed to the timing jitter of the SNSPD and MCP, resulting in a convoluted temporal width of 310 ps. The most straightforward solution is to replace the current TDC with a high-speed counterpart that offers a few picoseconds of resolution \cite{becker2021bh}, since the second-order correlation function fundamentally constructs both spectra. Data on the IRFs of the detectors and further discussion on the trigger choice of the idler spectrum are provided in the \href{Supplementary link}{Supplementary Information}.
\newline
\indent
\begin{figure*}[!t]
\begin{center}
\includegraphics[width=\linewidth]{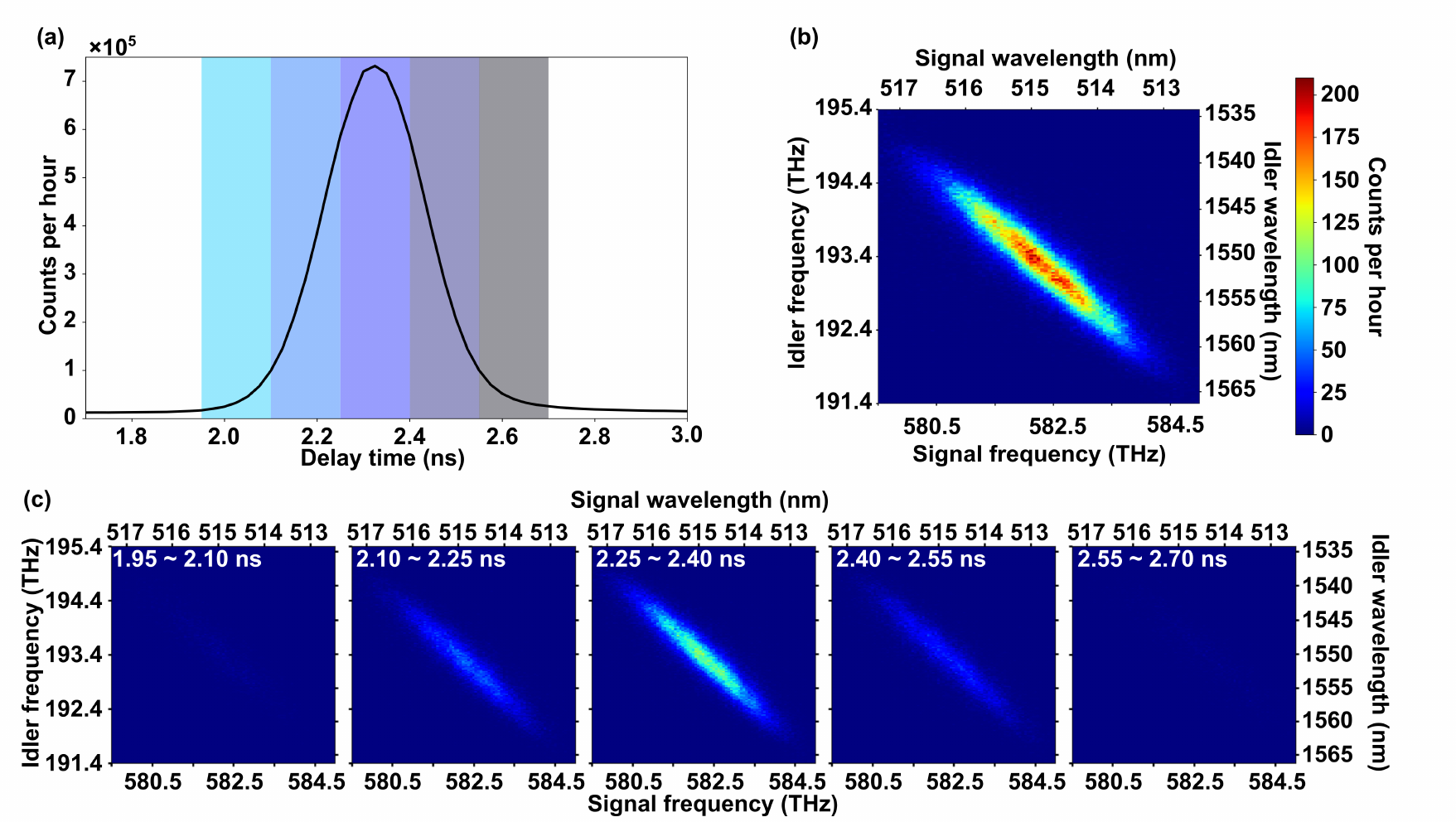}
\caption{(a) Instrument response function (IRF) measured between the MCP and the photodiode. (b) Non-time-resolved JSI includes the whole counts over the IRF. (c) The segmented JSIs are every 150 ps, with a range indicated on the top left, which corresponds to the segmented area of the IRF.}
\label{Main_Figure4}
\vspace{-7mm}
\end{center}
\end{figure*}
Figure \ref{Main_Figure4} illustrates both static and time-resolved JSIs. It should be noted that the photon counts of a JSI are dispersed over multiple JSI peaks, appearing every 13 ns because of the laser repetition rate. To avoid the loss of the photon count, the raw JSI data are folded every 13 ns using a modulo operation (see \href{Supplementary link}{Supplementary Information}). Fig. \ref{Main_Figure4}(a) is the IRF of the MCP triggered by the photodiode, where the temporal width of the peak is estimated to be 256 ps. The five coloured sections represent an equally segmented time window for creating JSI frames. The window width can vary from 25 ps upwards, limited by the resolution of the TDC. In this demonstration, the segmentation time is set to 150 ps, considering the contrast change of two-photon correlation on time-resolved JSIs. It should be noted that the system temporal resolution is 256 ps as indicated by the IRF, but the window value of 150 ps is set to observe the intensity change on the JSI. To align time-resolved JSIs with the static counterpart, the latter is shown in Fig. \ref{Main_Figure4}(b), with a maximum count of around 200. Here, we focus on the centre peak and the side lobes are ignored. Figure \ref{Main_Figure4}(c) is the array of time-resolved JSI, mapping the photon counts within the segmented time window of the IRF. It is clear that the counts in the frequency domain emerge within the first and second time windows, reaching the highest contrast in the third segment,  before vanishing across the fourth and fifth segments. The two-photon correlation is maintained across the five time windows, and contrast is solely affected by time segmentation. Photon count conservation is also confirmed between time-resolved and unresolved JSIs as the sum of the resolved counts on a pixel equals the unresolved count. These data provide the first proof of principle for our hybrid biphoton spectrometer, opening a new experimental avenue for integration of quantum information science and spectroscopy.
\section{Discussion}\label{sec3}
\begin{table*}[t]
\centering
\caption{Comparison of the spectroscopic systems for JSI measurements}
\resizebox{\textwidth}{!}{
\begin{tabular}{ccccccccc}\hline
References& \shortstack{Signal\\ wavelength (nm)}& \shortstack{Idler\\ wavelength (nm)}& \shortstack{Spectral\\ resolution (nm)\\ (Signal, Idler)} &System & Biphoton source& Scanning? & Non-degenerate? &Time-resolved?\\ 
\hline
\cite{PhysRevA.80.033814} & $\sim$789 & $\sim$633 & N/A & Dual grating spectrometer & SPDC & Yes & Yes & No\\
\cite{PhysRevA.81.031801} & $\sim$514 & $\sim$1547& (0.7, 24) & Dual prism spectrometer & SFWM & Yes & Yes & No\\
\cite{Chen:09} & $\sim$900 & $\sim$1300& (0.17, 0.33) & Tunable bandpass filters& SPDC &Yes & Yes  & No\\
\cite{wasilewski2006joint} & $\sim$720 & $\sim$850& N/A & Fourier spectroscopy & SPDC &Yes & Yes  & No\\
\cite{Fang:14} & $\sim$626 & $\sim$820& (0.06, 0.1) & Stimulated emission tomography & SFWM& Yes & Yes  & No\\
\cite{avenhaus2009fiber} & $\sim$1544 & $\sim$1517& (0.72, 0.72) & Dual fibre spectrometer& SPDC & No & Yes & No\\
\cite{lubin2021heralded} &$\sim$620 &$\sim$620& (2, 2) & Single SPAD array & Cascade emission & No & No & No\\
\cite{farella2024spectral} &$\sim$810 &$\sim$810& (0.52, 0.52) & Single data-driven camera & SPDC& No & No & No\\
This work& $\sim$515 & $\sim$1550& (0.63, 0.56) & Single fibre + DLD& SPDC & No & Yes & Yes\\
\hline
\end{tabular}}
\label{Comparison_JSI}
\vspace{-3mm}
\end{table*}
Table \ref{Comparison_JSI} provides a comprehensive comparison of spectroscopic systems for JSI measurements.
Scanning systems such as dual-grating \cite{PhysRevA.80.033814} or dual-prism \cite{PhysRevA.81.031801} spectrometers, or a set of tunable bandpass filters \cite{Chen:09} are straightforward ways to measure JSIs with both SPDC and spontaneous four-wave mixing (SFWM). Fourier spectroscopy can also construct JSI by measuring coincidence interferograms between two interferometric arms \cite{wasilewski2006joint}.
Although effective, these methods often suffer from low signal contrast and require prohibitively long acquisition times.
An alternative, stimulated emission tomography, can overcome these limitations, enabling highly resolved JSI measurements \cite{Fang:14}.
However, it still requires a chromatic laser scan to induce stimulated emission \cite{PhysRevLett.111.193602}. To circumvent the need for scanning, fibre spectrometers can be used \cite{avenhaus2009fiber}.
While successful in JSI measurements, their application is often limited to specific spectral ranges due to optical losses in the fibre \cite{Saleh:1084451}. Single-photon cameras, such as single-photon avalanche diode (SPAD) arrays \cite{lubin2021heralded} or data-driven cameras \cite{farella2024spectral}, offer another promising approach to non-scanning JSI acquisition. Crucially, none of the existing methodologies listed in Table \ref{Comparison_JSI} have demonstrated the ability to achieve highly non-degenerate time-resolved JSI measurements without scanning.
\newline
\indent
By combining a visible-sensitive DLD and a near-infrared fibre spectrometer, our approach uniquely enables the acquisition of highly non-degenerate JSI with time-resolving functionality relative to laser synchronisation, and eliminates the need for any scanning procedure. This advantage can be explained using the number of pixels. For instance, if we measure highly non-degenerate time-resolved JSIs with a $100 \times 100$ pixel array, the scanning system requires 10,000 scanning points, each of which accumulates photon counts for a certain time, increasing the measurement time. Furthermore, time-resolved JSI requires this scanning cycle several times. Although the dual fibre spectrometer can measure photons detected at all pixels without scanning, time-resolving a JSI relative to laser synchronisation is impossible because that temporal information is already used to measure the spectra themselves. In terms of single-photon-sensitive cameras, they can obtain counts from 10,000 pixels without scanning, and we cannot explicitly assess whether they can measure time-resolved JSI, as we have no experience with them. However, it would be challenging to process time tags generated from 10,000 pixels and synchronise them with the laser signal. If such an operation is possible, the degenerate time-resolved JSI could be obtained with a single camera. Compared with the expected complexity, our hybrid system requires only five streams of time tags to reconstruct time-resolved JSIs.
\newline
\indent
The capability for time-resolved JSI measurements opens up unprecedented opportunities in various fields of spectroscopy, including single-molecule spectroscopy \cite{moerner2003methods}, Raman spectroscopy \cite{orlando2021comprehensive}, two-dimensional infrared spectroscopy \cite{Hamm_Zanni_2011}, or nuclear magnetic resonance spectroscopy \cite{gunther2013nmr}.
Our system is particularly well-suited for time-resolved fluorescence spectroscopy excited by heralded single photons \cite{fujihashi2020generation,gabler2025benchmarking, doi:10.1021/acs.jpclett.3c01266,Eshun:23, tsao2025heralded,li2025comparing}, where one photon from an entangled pair excites the sample and the other acts as a temporal herald.
This new capability could pave the way for investigating complex biological, chemical, and physical systems.
In fact, single-photon fluorescence spectroscopy has already been used to study natural photosynthetic complexes, probing energy transfer between chromophores under near-sunlight illumination conditions \cite{li2023single}.
\newline
\indent
Building upon this, a novel scheme for two-dimensional fluorescence spectroscopy (2DFS) is developed using entangled photons \cite{fujihashi2026two, fujihashi2026two_JCP}.
This theoretical simulation revealed the compatibility of two-dimensional electronic spectroscopy (2DES) \cite{fresch2023two}, often used to elucidate complex molecular systems in ultrafast laser spectroscopy.
This novel quantum spectroscopic scheme offers significant advantages over conventional 2DES by leveraging the non-local frequency correlations of entangled photon pairs.
This results in a reduction of the spectral complexity in 2DES and could provide a deeper insight into the physical function of complex molecular systems.
The hybrid spectrometer presented here serves as a direct platform for 2DFS with time-frequency entangled photons, which requires
time- and frequency-resolved measurements of both excitation and fluorescence photons.
\newline
\indent
While our current temporal resolution of a few hundred picoseconds is a significant step, it is still insufficient to directly observe single-photon wave packets \cite{kuzucu2008joint, yabuno2022ultrafast}, which range from a few picoseconds to a hundred femtoseconds, as well as quantum coherence in photosynthetic systems \cite{engel2007evidence}. Even with these limitations, our system is sufficient to measure some dynamical processes in molecules excited by heralded single photons, such as photosynthetic complexes \cite{alvarez2025entangled}.
A clear improvement of temporal resolution involves integrating a streak tube \cite{wiersig2009direct} with the unused y-coordinate of the DLD. This enhancement would enable the investigation of crucial molecular dynamics in systems in light-harvesting complexes \cite{li2023single, alvarez2025entangled}, fluorescent proteins \cite{kim2019venusa206, feder2025fluorescent}, or organic semiconductors \cite{uoyama2012highly,fairbairn2026highly}.
In essence, our methodology not only provides a new tool for quantum spectroscopy but also offers a clear and powerful framework for its future evolution. 
\newline
\indent
In conclusion, we have demonstrated a hybrid biphoton spectrometer for visible and near-infrared photon pairs generated from the SPDC crystal. The individual one-dimensional spectra are first measured with the delay-line-anode single-photon detector for the signal photons and the fibre spectrometer for the idler photons. Then, static JSI is reconstructed by processing time tags from two spectrometers, yielding the highly non-degenerate JSI. The time-resolved JSI is reconstructed solely by measuring the temporal information of signal photons, and the photon counts on the JSI vary according to the defined time window. Because it requires no scanning and provides time-resolved capability, this work lays the foundation for a new generation of quantum spectroscopic techniques that could revolutionise our understanding of complex molecular and biological dynamics.

\begin{backmatter}
\bmsection{Funding} 
MEXT Quantum Leap Flagship Program (MEXT Q-LEAP, JPMXS0118069242); Japan Society for the Promotion of Science (JSPS KAKENHI, 25KF0034).
\bmsection{Acknowledgment} 
We are grateful to Achim Czasch in RoentDek Handels GmbH for their support in TDC data processing for the delay-line-anode single-photon detector. We acknowledge fruitful discussions with Gordon J. Hedley regarding the future application.
\bmsection{Disclosures} 
The authors declare no conflicts of interest.
\bmsection{Data Availability} 
See \hyperlink{blank_url}{Supplementary information} for supporting content.
\end{backmatter}

\bibliography{sample}

@article{li2023single,
  title={Single-photon absorption and emission from a natural photosynthetic complex},
  author={Li, Quanwei and Orcutt, Kaydren and Cook, Robert L and Sabines-Chesterking, Javier and Tong, Ashley L and Schlau-Cohen, Gabriela S and Zhang, Xiang and Fleming, Graham R and Whaley, K Birgitta},
  journal={Nature},
  volume={619},
  number={7969},
  pages={300--304},
  year={2023},
  publisher={Nature Publishing Group UK London}
}

@article{doi:10.1021/acs.jpclett.3c01266,
author = {Harper, Nathan and Hickam, Bryce P. and He, Manni and Cushing, Scott K.},
title = {Entangled Photon Correlations Allow a Continuous-Wave Laser Diode to Measure Single-Photon, Time-Resolved Fluorescence},
journal = {J. Phys. Chem. Lett.},
volume = {14},
number = {25},
pages = {5805-5811},
year = {2023},
}

@article{Eshun:23,
author = {Audrey Eshun and Xiyu Yi and Ashleigh Wilson and Sam Jeppson and Jae Hyuck Yoo and Shervin Kiannejad and Mike Rushford and Tiziana Bond and Ted Laurence},
journal = {Opt. Express},
number = {16},
pages = {26935--26947},
publisher = {Optica Publishing Group},
title = {Fluorescence lifetime measurements using photon pair correlations generated via spontaneous parametric down conversion ({SPDC})},
volume = {31},
month = {Jul},
year = {2023},
}

@article{gabler2025benchmarking,
  title={Benchmarking of fluorescence lifetime measurements using time-frequency correlated photons},
  author={G{\"a}bler, Tobias B and Then, Patrick and Eggeling, Christian and Gr{\"a}fe, Markus and Jain, Nitish and Gili, Valerio F},
  journal={Methods in Microscopy},
  volume={2},
  number={1},
  pages={133--145},
  year={2025},
  publisher={De Gruyter}
}

@article{fujihashi2026two,
  title={Two-dimensional fluorescence spectroscopy with entangled photons and time-and frequency-resolved two-photon coincidence detection},
  author={Fujihashi, Yuta and Iso, Ozora and Shimizu, Ryosuke and Ishizaki, Akihito},
  journal={Science Advances},
  volume={12},
  number={16},
  pages={eaed7026},
  year={2026},
  publisher={American Association for the Advancement of Science}
}

@article{Iso:25,
author = {Ozora Iso and Kensuke Miyajima and Ryosuke Shimizu},
journal = {Opt. Express},
number = {9},
pages = {19504--19513},
publisher = {Optica Publishing Group},
title = {Capturing the spectrotemporal structure of a biphoton wave packet with delay-line-anode single-photon imagers},
volume = {33},
month = {May},
year = {2025},
url = {https://opg.optica.org/oe/abstract.cfm?URI=oe-33-9-19504},
doi = {10.1364/OE.537512},
}

@article{PhysRevA.80.033814,
  title = {Spectral properties of entangled photons generated via type-{I} frequency-nondegenerate spontaneous parametric down-conversion},
  author = {Baek, So-Young and Kim, Yoon-Ho},
  journal = {Phys. Rev. A},
  volume = {80},
  issue = {3},
  pages = {033814},
  numpages = {8},
  year = {2009},
  month = {Sep},
  publisher = {American Physical Society},
}

@article{Chen:09,
author = {Jun Chen and Aaron J. Pearlman and Alexander Ling and Jingyun Fan and Alan Migdall},
journal = {Opt. Express},
number = {8},
pages = {6727--6740},
publisher = {Optica Publishing Group},
title = {A versatile waveguide source of photon pairs for chip-scale quantum information processing},
volume = {17},
month = {Apr},
year = {2009},
}

@article{PhysRevA.81.031801,
  title = {Bridging visible and telecom wavelengths with a single-mode broadband photon pair source},
  author = {S\"oller, C. and Brecht, B. and Mosley, P. J. and Zang, L. Y. and Podlipensky, A. and Joly, N. Y. and Russell, P. St. J. and Silberhorn, C.},
  journal = {Phys. Rev. A},
  volume = {81},
  issue = {3},
  pages = {031801},
  numpages = {4},
  year = {2010},
  month = {Mar},
  publisher = {American Physical Society},
}

@article{Fang:14,
author = {Bin Fang and Offir Cohen and Marco Liscidini and John E. Sipe and Virginia O. Lorenz},
journal = {Optica},
number = {5},
pages = {281--284},
publisher = {Optica Publishing Group},
title = {Fast and highly resolved capture of the joint spectral density of photon pairs},
volume = {1},
month = {Nov},
year = {2014},
}

@article{PhysRevLett.111.193602,
  title = {Stimulated Emission Tomography},
  author = {Liscidini, M. and Sipe, J. E.},
  journal = {Phys. Rev. Lett.},
  volume = {111},
  issue = {19},
  pages = {193602},
  numpages = {5},
  year = {2013},
  month = {Nov},
  publisher = {American Physical Society},
}

@book{Hamm_Zanni_2011,
  title     = {Concepts and Methods of 2D Infrared Spectroscopy},
  author    = {Hamm, Peter and Zanni, Martin},
  year      = {2011},
  publisher = {Cambridge University Press},
  address   = {Cambridge},
}

@article{JAGUTZKI2002244,
title = {A broad-application microchannel-plate detector system for advanced particle or photon detection tasks: large area imaging, precise multi-hit timing information and high detection rate},
journal = {Nuclear Instruments and Methods in Physics Research Section A: Accelerators, Spectrometers, Detectors and Associated Equipment},
volume = {477},
number = {1},
pages = {244-249},
year = {2002},
note = {5th Int. Conf. on Position-Sensitive Detectors},
issn = {0168-9002},
author = {O Jagutzki and V Mergel and K Ullmann-Pfleger and L Spielberger and U Spillmann and R Dörner and H Schmidt-Böcking},
}

@book{Saleh:1084451,
      author        = "Saleh, Bahaa E A and Teich, Malvin Carl",
      title         = "{Fundamentals of photonics; 2nd ed.}",
      publisher     = "Wiley",
      address       = "New York, NY",
      series        = "Wiley series in pure and applied optics",
      year          = "2007",
}

@article{alvarez2025entangled,
  author = {Álvarez-Mendoza, Raúl and Uboldi, Lorenzo and Lyons, Ashley and Cogdell, Richard J. and Cerullo, Giulio and Faccio, Daniele},
  title        = {Correlated-photon time- and frequency-resolved optical spectroscopy},
  journal      = {Nature Communications},
  volume       = {16},
  pages        = {8634},
  year         = {2025},
}

@article{Zielnicki07062018,
author = {Kevin Zielnicki and Karina Garay-Palmett and Daniel Cruz-Delgado and Hector Cruz-Ramirez and Michael F. O’Boyle and Bin Fang and Virginia O. Lorenz and Alfred B. U’Ren and Paul G. Kwiat and},
title = {Joint spectral characterization of photon-pair sources},
journal = {Journal of Modern Optics},
volume = {65},
number = {10},
pages = {1141--1160},
year = {2018},
publisher = {Taylor \& Francis},
}

@article{avenhaus2009fiber,
  title={Fiber-assisted single-photon spectrograph},
  author={Avenhaus, Malte and Eckstein, Andreas and Mosley, Peter J and Silberhorn, Christine},
  journal={Optics Letters},
  volume={34},
  number={18},
  pages={2873--2875},
  year={2009},
  publisher={Optical Society of America}
}

@article{farella2024spectral,
    author = {Farella, Brianna and Medwig, Gregory and Abrahao, Raphael A. and Nomerotski, Andrei},
    title = {Spectral characterization of an {SPDC} source with a fast broadband spectrometer},
    journal = {AIP Advances},
    volume = {14},
    number = {4},
    pages = {045034},
    year = {2024},
    month = {04},
    issn = {2158-3226},
    doi = {10.1063/5.0168423},
}

@article{lubin2021heralded,
  title={Heralded spectroscopy reveals exciton--exciton correlations in single colloidal quantum dots},
  author={Lubin, Gur and Tenne, Ron and Ulku, Arin Can and Antolovic, Ivan Michel and Burri, Samuel and Karg, Sean and Yallapragada, Venkata Jayasurya and Bruschini, Claudio and Charbon, Edoardo and Oron, Dan},
  journal={Nano Letters},
  volume={21},
  number={16},
  pages={6756--6763},
  year={2021},
  publisher={ACS Publications}
}

@article{moerner2003methods,
  title={Methods of single-molecule fluorescence spectroscopy and microscopy},
  author={Moerner, WE and Fromm, David P},
  journal={Rev. Sci. Instrum.},
  volume={74},
  number={8},
  pages={3597--3619},
  year={2003},
  publisher={American Institute of Physics}
}

@article{orlando2021comprehensive,
  title={A comprehensive review on Raman spectroscopy applications},
  author={Orlando, Andrea and Franceschini, Filippo and Muscas, Cristian and Pidkova, Solomiya and Bartoli, Mattia and Rovere, Massimo and Tagliaferro, Alberto},
  journal={Chemosensors},
  volume={9},
  number={9},
  pages={262},
  year={2021},
  publisher={MDPI}
}

@book{becker2021bh,
  author    = {Becker, Wolfgang},
  title     = {The bh TCSPC Handbook},
  publisher = {Becker \& Hickl GmbH},
  year      = {2021},
  address   = {Berlin},
  edition   = {8th},
}

@article{kim2019venusa206,
  title={Venus\textsubscript{A206} dimers behave coherently at room temperature},
  author={Kim, Youngchan and Puhl, Henry L and Chen, Eefei and Taumoefolau, Grace H and Nguyen, Tuan A and Kliger, David S and Blank, Paul S and Vogel, Steven S},
  journal={Biophysical Journal},
  volume={116},
  number={10},
  pages={1918--1930},
  year={2019},
  publisher={Elsevier}
}

@article{wiersig2009direct,
  title={Direct observation of correlations between individual photon emission events of a microcavity laser},
  author={Wiersig, J and Gies, C and Jahnke, F and A{\ss}mann, M and Berstermann, T and Bayer, M and Kistner, C and Reitzenstein, S and Schneider, Ch and H{\"o}fling, Sven and others},
  journal={Nature},
  volume={460},
  number={7252},
  pages={245--249},
  year={2009},
  publisher={Nature Publishing Group UK London}
}

@article{ladd2010quantum,
  title={Quantum computers},
  author={Ladd, Thaddeus D and Jelezko, Fedor and Laflamme, Raymond and Nakamura, Yasunobu and Monroe, Christopher and O’Brien, Jeremy Lloyd},
  journal={Nature},
  volume={464},
  number={7285},
  pages={45--53},
  year={2010},
  publisher={Nature Publishing Group UK London}
}

@article{gisin2007quantum,
  title={Quantum communication},
  author={Gisin, Nicolas and Thew, Rob},
  journal={Nature Photonics},
  volume={1},
  number={3},
  pages={165--171},
  year={2007},
  publisher={Nature Publishing Group UK London}
}

@article{toth2014quantum,
  title={Quantum metrology from a quantum information science perspective},
  author={T{\'o}th, G{\'e}za and Apellaniz, Iagoba},
  journal={Journal of Physics A: Mathematical and Theoretical},
  volume={47},
  number={42},
  pages={424006},
  year={2014},
  publisher={IOP Publishing}
}

@book{gunther2013nmr,
  title     = {NMR Spectroscopy: Basic Principles, Concepts and Applications in Chemistry},
  author    = {G{\"u}nther, Harald},
  year      = {2013},
  edition   = {3rd},
  publisher = {John Wiley \& Sons},
  address   = {Weinheim},
}

@article{engel2007evidence,
  title={Evidence for wavelike energy transfer through quantum coherence in photosynthetic systems},
  author={Engel, Gregory S and Calhoun, Tessa R and Read, Elizabeth L and Ahn, Tae-Kyu and Man{\v{c}}al, Tom{\'a}{\v{s}} and Cheng, Yuan-Chung and Blankenship, Robert E and Fleming, Graham R},
  journal={Nature},
  volume={446},
  number={7137},
  pages={782--786},
  year={2007},
  publisher={Nature Publishing Group UK London}
}

@article{feder2025fluorescent,
  title={A fluorescent-protein spin qubit},
  author={Feder, Jacob S and Soloway, Benjamin S and Verma, Shreya and Geng, Zhi Z and Wang, Shihao and Kifle, Bethel B and Riendeau, Emmeline G and Tsaturyan, Yeghishe and Weiss, Leah R and Xie, Mouzhe and others},
  journal={Nature},
  volume={645},
  number={8079},
  pages={73--79},
  year={2025},
  publisher={Nature Publishing Group UK London}
}

@article{uoyama2012highly,
  title={Highly efficient organic light-emitting diodes from delayed fluorescence},
  author={Uoyama, Hiroki and Goushi, Kenichi and Shizu, Katsuyuki and Nomura, Hiroko and Adachi, Chihaya},
  journal={Nature},
  volume={492},
  number={7428},
  pages={234--238},
  year={2012},
  publisher={Nature Publishing Group UK London}
}

@article{fairbairn2026highly,
  author    = {Fairbairn, N. J. and Vodianova, O. and Schmidt, B. V. K. J. and others},
  title     = {Highly efficient exciton-exciton annihilation in single conjugated polymer chains},
  journal   = {Nature Communications},
  volume    = {17},
  pages     = {731},
  year      = {2026},
  doi       = {10.1038/s41467-025-67422-z}
}

@article{tsao2025heralded,
title={Heralded Emission Detection in {InAs/ZnSe} Quantum Dot Solids Using Time-Correlated Photons},
author={Tsao, Chieh and Li, Xiang and Hinkle, Alex and Chen, Yifan and Oskarsson, Elvar and Banin, Uri and Utzat, Hendrik},
journal={arXiv preprint arXiv:2509.11704},
year={2025} 
}

@article{gong2022attosecond,
title={Attosecond spectroscopy of size-resolved water clusters},
author={Gong, Xiaochun and Heck, Saijoscha and Jelovina, Denis and Perry, Conaill and Zinchenko, Kristina and Lucchese, Robert and W{\"o}rner, Hans Jakob},
journal={Nature},
volume={609}, 
number={7927},
pages={507--511},
year={2022},
publisher={Nature Publishing Group UK London} }

@article{li2025imaging, title={Imaging a light-induced molecular elimination reaction with an {X}-ray free-electron laser},
author={Li, Xiang and Boll, Rebecca and Vindel-Zandbergen, Patricia and Gonz{\'a}lez-V{\'a}zquez, Jes{\'u}s and Rivas, Daniel E and Bhattacharyya, Surjendu and Borne, Kurtis and Chen, Keyu and De Fanis, Alberto and Erk, Benjamin and others},
journal={Nature Communications}, 
volume={16},
number={1},
pages={7006},
year={2025},
publisher={Nature Publishing Group UK London} }

@article{schnorr2014electron,
  title =  {Electron Rearrangement Dynamics in Dissociating {${\mathrm{I}}_{2}^{n+}$} Molecules Accessed by Extreme Ultraviolet Pump-Probe Experiments},
  author = {Schnorr, K. and Senftleben, A. and Kurka, M. and Rudenko, A. and Schmid, G. and Pfeifer, T. and Meyer, K. and K\"ubel, M. and Kling, M. F. and Jiang, Y. H. and Treusch, R. and D\"usterer, S. and Siemer, B. and W\"ostmann, M. and Zacharias, H. and Mitzner, R. and Zouros, T. J. M. and Ullrich, J. and Schr\"oter, C. D. and Moshammer, R.},
  journal = {Phys. Rev. Lett.},
  volume = {113},
  issue = {7},
  pages = {073001},
  numpages = {5},
  year = {2014},
  month = {Aug},
  publisher = {American Physical Society},
}

@article{maclean2018direct,
title={Direct characterization of ultrafast energy-time entangled photon pairs},
author={MacLean, Jean-Philippe W and Donohue, John M and Resch, Kevin J},
journal={Phys. Rev. Lett.},
volume={120},
number={5},
pages={053601},
year={2018},
publisher={APS} }

@article{davis2020measuring,
title={Measuring the quantum state of a photon pair entangled in frequency and time},
author={Davis, Alex OC and Thiel, Val{\'e}rian and Smith, Brian J}, 
journal={Optica},
volume={7},
number={10},
pages={1317--1322},
year={2020},
publisher={Optical Society of America} }

@article{kuzucu2008joint,
title={Joint temporal density measurements for two-photon state characterization},
author={Kuzucu, Onur and Wong, Franco NC and Kurimura, Sunao and Tovstonog, Sergey},
journal={Phys. Rev. Lett.},
volume={101},
number={15},
pages={153602},
year={2008},
publisher={APS} }

@article{yabuno2022ultrafast,
title={Ultrafast measurement of a single-photon wave packet using an optical {K}err gate},
author={Yabuno, Masahiro and Takumi, Takahiro and China, Fumihiro and Miki, Shigehito and Terai, Hirotaka and Mosley, Peter J and Jin, Rui-Bo and Shimizu, Ryosuke},
journal={Optics Express},
volume={30},
number={4},
pages={4999--5007},
year={2022},
publisher={Optica Publishing Group} }

@article{sansonetti1996wavelengths,
title={Wavelengths of spectral lines in mercury pencil lamps},
author={Sansonetti, Craig J and Salit, Marc L and Reader, Joseph},
journal={Applied Optics}, 
volume={35},
number={1},
pages={74--77}, 
year={1996},
publisher={Optical Society of America} }

@article{wasilewski2006joint,
title={Joint spectrum of photon pairs measured by coincidence {F}ourier spectroscopy},
author={Wasilewski, Wojciech and Wasylczyk, Piotr and Kolenderski, Piotr and Banaszek, Konrad and Radzewicz, Czes{\l}aw},
journal={Optics Letters},
volume={31}, number={8},
pages={1130--1132},
year={2006},
publisher={Optical Society of America} }

@article{jin2018time,
title={Time-frequency duality of biphotons for quantum optical synthesis},
author={Jin, Rui-Bo and Saito, Takuma and Shimizu, Ryosuke},
journal={Physical Review Applied},
volume={10},
number={3},
pages={034011},
year={2018},
publisher={APS} }

@article{Kuwana_2026,
doi = {10.35848/1347-4065/ae7da9},
url = {https://doi.org/10.35848/1347-4065/ae7da9},
year = {2026},
month = {jul},
publisher = {IOP Publishing},
volume = {65},
number = {13},
pages = {130901},
author = {Kuwana, Takahisa and Yabuno, Masahiro and Terai, Hirotaka and Miki, Shigehito and Mosley, Peter J. and Jin, Rui-Bo and Shimizu, Ryosuke},
title = {Evaluating the resolution-efficiency trade-off of optical {K}err gating via biphoton temporal correlation measurements},
journal = {Japanese Journal of Applied Physics},
}

@article{fresch2023two,
title={Two-dimensional electronic spectroscopy},
author={Fresch, Elisa and Camargo, Franco VA and Shen, Qijie and Bellora, Caitlin C and Pullerits, T{\~o}nu and Engel, Gregory S and Cerullo, Giulio and Collini, Elisabetta},
journal={Nature Reviews Methods Primers},
volume={3},
number={1},
pages={84},
year={2023},
publisher={Nature Publishing Group UK London}
}

@article{czasch2007position,
title={Position- and time-sensitive single photon detector with delay-line readout},
author={Czasch, A and Milnes, J and Hay, N and Wicking, W and Jagutzki, O},
journal={Nuclear Instruments and Methods in Physics Research Section A: Accelerators, Spectrometers, Detectors and Associated Equipment},
volume={580},
number={2},
pages={1066--1070},
year={2007},
publisher={Elsevier} }

@article{fujihashi2026two_JCP,
    author = {Fujihashi, Yuta and Ishizaki, Akihito},
    title = {Two-dimensional fluorescence spectroscopy with quantum entangled photons: Idler-referenced timing without pump detection},
    journal = {J. Chem. Phys.},
    volume = {164},
    number = {22},
    pages = {224108},
    year = {2026},
    month = {06},
    issn = {0021-9606},
}

@article{fujihashi2020generation,
title={Generation of pseudo-sunlight via quantum entangled photons and the interaction with molecules},
author={Fujihashi, Yuta and Shimizu, Ryosuke and Ishizaki, Akihito},
journal={Physical Review Research},   volume={2},
number={2},
pages={023256},
year={2020},
publisher={APS} 
}

@article{li2025comparing,
title={Comparing photosynthetic light harvesting of single photons and pseudothermal light under ultraweak illumination},
author={Li, Quanwei and Ko, Liwen and Whaley, K Birgitta and Fleming, Graham R},
journal={Science Advances},
volume={11},
number={46},
pages={eadz2616},
year={2025},
publisher={American Association for the Advancement of Science} 
}

\bibliographyfullrefs{sample}

\ifthenelse{\equal{\journalref}{aop}}{%
\section*{Author Biographies}
\begingroup
\setlength\intextsep{0pt}
\begin{minipage}[t][6.3cm][t]{1.0\textwidth} 
  \begin{wrapfigure}{L}{0.25\textwidth}
    \includegraphics[width=0.25\textwidth]{john_smith.eps}
  \end{wrapfigure}
  \noindent
  {\bfseries John Smith} received his BSc (Mathematics) in 2000 from The University of Maryland. His research interests include lasers and optics.
\end{minipage}
\begin{minipage}{1.0\textwidth}
  \begin{wrapfigure}{L}{0.25\textwidth}
    \includegraphics[width=0.25\textwidth]{alice_smith.eps}
  \end{wrapfigure}
  \noindent
  {\bfseries Alice Smith} also received her BSc (Mathematics) in 2000 from The University of Maryland. Her research interests also include lasers and optics.
\end{minipage}
\endgroup
}{}

\end{document}